\input ./style/arxiv-general.cfg
\documentclass[aoas,MSNbibl,nameyear,rotating,dvips]{arximspdf}
\makeatletter
   \@ifpackageloaded{graphicx}{}{\usepackage{graphicx}}
\makeatother

\doi{10.1214/15-AOAS811}
\volume{9}
\issue{2}
\pubyear{2015}
\firstpage{597}
\lastpage{620}
\docsubty{FLA}

\makeatletter
\newcommand{\rrvert}{\vert}
\newcommand{\llvert}{\vert}
\renewcommand{\epsilon}{\varepsilon}
\newproclaim{remark}{Remark}

\makeatother

\begin{document}
\begin{frontmatter}

\title{Regression based principal component analysis for sparse functional data with applications to screening growth paths}
\runtitle{Regression based principal component analysis}

\begin{aug}
\author[A]{\fnms{Wenfei}~\snm{Zhang}\corref{}\ead[label=e11]{wenfeizhang@gmail.com}} \and
\author[A]{\fnms{Ying}~\snm{Wei}\ead[label=e1]{ying.wei@columbia.edu}}
\runauthor{W. Zhang and Y. Wei}
\affiliation{Columbia University}
\address[A]{Department of Biostatistics\\
Columbia University\\
722 West 168th St. Rm 644\\
New York, New York 10032\\
USA\\
\printead{e1}}
\end{aug}

%
\received{\smonth{5} \syear{2014}}
%
\revised{\smonth{1} \syear{2015}}

\begin{abstract}
Growth charts are widely used in pediatric care for assessing childhood
body size measurements (e.g., height or weight). The existing growth
charts screen one body size at a single given age. However, when a
child has multiple measures over time and exhibits a growth path, how
to assess those measures jointly in a rigorous and quantitative way
remains largely undeveloped in the literature. In this paper, we
develop a new method to construct growth charts for growth paths. A new
estimation algorithm using alternating regressions is developed to
obtain principal component representations of growth paths (sparse
functional data). The new algorithm does not rely on strong
distribution assumptions and is computationally robust and easily
incorporates subject level covariates, such as parental information.
Simulation studies are conducted to investigate the performance of our
proposed method, including comparisons to existing methods. When the
proposed method is applied to monitor the puberty growth among a group
of Finnish teens, it yields interesting insights.
%
\end{abstract}

\begin{keyword}
\kwd{Growth charts}
\kwd{sparse functional data}
\kwd{longitudinal data}
\kwd{principal component analysis}
\end{keyword}
\end{frontmatter}

\section{Introduction}\label{SecIntro}
In pediatric practice, height, weight and other body size measurements
are frequently examined for infants, children and adolescents in order
to ensure their healthy growth. The most commonly used tools are growth
charts, also known as reference centile charts. The fundamental purpose
of growth charts is to identify percentile ranks of individuals with
respect to their corresponding reference populations, and to screen out
subjects with extreme ranks, either too high or too low, for further
medical investigations. The conventional growth charts consist of a
series of percentile curves for a certain measurement over ages. Those
percentile curves are estimated from a reference population using
penalized likelihood methods introduced in \citet{Cole1988} and
\citet{Colegreen1992}. They are used to identify individual percentile ranks
at specific ages. Lately, several methods, including \citet{Thompson1997},
\citet{Scheike1999}, \citet{Wei2006} and
\citet{chen2012conditional}, were proposed to further incorporate prior
information and subject level covariates into growth charts. In these
methods, the reference percentiles are estimated by conditioning on not
only target ages but also prior measurements and other important
variables, such as prognostic and parental information. The resulting
growth charts are hence called conditional growth charts.
\citet{Thompson1997} assumed a multivariate normal distribution for the
measurements and the covariates at all time points and used the maximum
likelihood estimator for the mean and variance functions.
\citet{Scheike1999} considered a longitudinal regression model accounting for
the previous measurement adjacent to the current measurement and the
duration in between. To avoid a particular distributional assumption,
\citet{Wei2006} proposed a semi-parametric quantile regression model to
construct conditional growth charts.

\begin{figure}[b]

\includegraphics{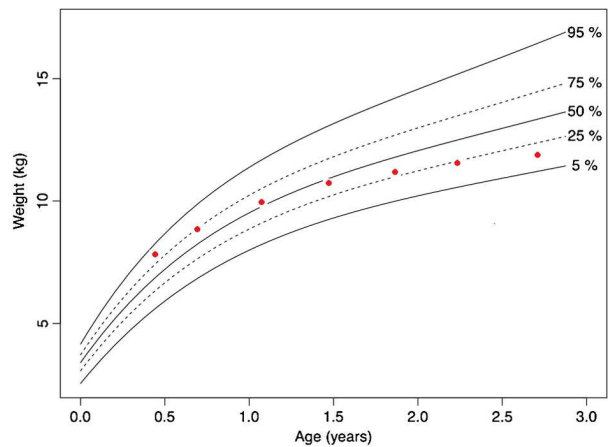}

\caption{An example of an abnormal growth pattern. The dots represent
the growth path for a subject. The curves are the percentile curves at
quantile levels 0.05, 0.25, 0.50, 0.75 and 0.95. The $x$ axis represents ages.
The $y$ axis represents\vspace*{-6pt} weights.}\label{F000growthchartexample1}
\end{figure}

Both conventional and conditional growth charts screen only one single
measurement at a time. However, due to common clinical practice, each
individual has its measurements collected longitudinally and exhibits a
growth path over time. A growth path may not be normal even if each of
its measurements is within the normal ranges of both conventional and
conditional growth charts. For example, as shown in Figure~\ref{F000growthchartexample1}, this subject starts at the 90th percentile
in weight at the age of~0.5, and gradually declines to the 15th
percentile around the age of 2.5. Although such a slow decline in the
growth path should be alerting, it cannot be recognized by conventional
growth charts, because all its measurements are within the normal
ranges. It cannot be detected by conditional growth charts either,
since the changes from the preceding measurements are not large enough.

Therefore, screening entire growth paths may bring new insights into
growth screening. However, existing screening methods for growth paths
are mostly empirical, relying heavily on personal experiences of
medical providers [\citet{Legler1998}]. Rigorous quantitative screening
methods for entire growth paths remain largely undeveloped. Hence, in
this paper, we propose a new statistical method to construct growth
charts that enable the screening of entire growth paths.

Growth charts are estimated from a reference growth data set, which is
collected from a representative sample in a target population, and
consists of longitudinal body size measurements. Most reference growth
data share the following characteristics. First, each growth path is
only observed at sparse and irregularly spaced time points with
possible measurement errors. Therefore, statistical tools developed for
multivariate and functional data are directly applicable, as the former
requires a fixed measurement schedule and the latter requires densely
observed data on each growth path. Second, the distributions of body
size measurements are unlikely to follow certain parametric
distributions. Therefore, likelihood based parametric approaches are
often undesirable in such applications.

Considering the characteristics of reference growth data, we develop a
two-step procedure for identifying percentile ranks of growth paths. In
the first step, we propose a novel regression based principal component
analysis (PCA) algorithm that is tailored specifically for reference
growth data. In the second step, we construct the multivariate quantile
contours of the resulting component scores, which can be used to
identify percentile ranks of growth paths. The proposed PCA algorithm
can also incorporate covariates, which in turn enables the screening of
growth paths conditioned on individual characteristics.

The rest of this paper is organized into the following structure: In
Section~\ref{Secscreeningmethod} we elaborate on the proposed screening
method, including the general model settings and notation in Section~\ref{settingNotation}, the introduction of the proposed regression
based principal component analysis in Section~\ref{PCAsparse},
the construction of growth charts for screening growth paths in Section~\ref{growthchartConstruction}, and the extension of incorporating
covariates in Section~\ref{covariateSec}. In Section~\ref{ChapterApp}
we provide examples of applying the proposed method in the field of
pediatrics. In Section~\ref{ChapterNumerical} we present the numerical
investigation of our method. In Section~\ref{conclusion} we include
discussions and conclusions on the important findings.

\section{Methods}\label{Secscreeningmethod}
\subsection{Settings and notation}\label{settingNotation}
A reference growth data set consists of $N$ subjects and their
longitudinal measurements $\{Y_{ij},T_{ij}\}_{i=1,\ldots, N,j=1,\ldots,
m_i}$. Here $m_i$ is the number of measurements for the $i$th subject,
and $Y_{ij}$ is the $j$th observation for the $i$th subject measured at
the time of $T_{ij}$, $T_{ij}\in\mathcal{T}$. We assume that each
longitudinal growth path is observed from the following model:
%
\begin{equation}
\label{equaYij} Y_{ij}=Y_i(T_{ij})+
\epsilon_{ij}, \qquad T_{ij}\in\mathcal{T},
\end{equation}
where\vspace*{1pt} $Y_i(t)$'s are the underlying growth paths, $\epsilon_{ij}$'s,
and independent of $Y_i(t)$'s, are i.i.d. random errors with mean zero
and constant variance $\sigma^2$. $\epsilon_{ij}$ can be viewed as the
measurement error associated with $Y_{ij}$, and we implicitly assume
that measurement errors do not depend on magnitudes of measurements and
measurement times. Such assumptions are reasonable for reference growth
data. For example, the weight measurement error due to a weight scale
is usually related neither to the weight itself nor to the time when
the weight is taken.

By the Karhunen--Lo\`{e}ve theorem in \citet{loeve1978}, the true growth
paths $Y_i(t)$, if smooth and continuous, can be written as
%
\begin{equation}
\label{Ydecompsition}
Y_i(t) = U(t)+\sum_{k=1}^\infty
r_{ik}\phi_k(t),
\end{equation}
where $U(t) = E\{Y_i(t)\}$ is the population mean function, $\phi
_k(t)$'s are principal
component functions, which are continuous pair-wise orthogonal
functions on $\mathcal{T}$ with $\int_{\mathcal{T}}\phi_k(t)^2\,dt=1$,
and $r_{ik}$'s are principal component scores, which are uncorrelated
random variables with mean 0 and variance $\lambda_k$, where $\lambda
_1\geq\lambda_2,\ldots.$
This decomposition provides the basis of PCA
for functional data.

We further assume $Y_i(t)$ can be well approximated by the first $K$
principal component functions, that is, $Y_i(t) \approx U(t)+\sum_{k=1}^Kr_{ik}\phi_k(t)$. This approximation is biologically plausible,
since the biological growth process is mainly driven by several growth
hormones, as mentioned in \citet{zhang2012regression}. As each growth
hormone determines a particular growth pattern, the observed growth
path is the result of their joint actions. Therefore, with the $k$th
component function $\phi_k$ representing a certain growth pattern, the
component score $r_{ik}$ measures the extent to which $\phi_k(t)$
contributes to the individual growth path $Y_i(t)$. The biological
meaning of component functions $\phi_k(t)$ and scores $r_{ik}$ is also
exemplified in Section~\ref{secApp}. This way, the distribution of the
growth paths, $Y_i(t)$'s, are fully determined by their component
scores. Consequently, the growth charts for $Y_i(t)$ can be constructed
based on the joint distribution of the first $K$ component scores. To
estimate the component functions of $Y_i(t)$ from the reference growth
data, we proposed a regression based PCA algorithm in Section~\ref{PCAsparse}.

The following notation will be used to illustrate our proposed method:
$L^2(\mathcal{T})$ is the set of square integrable functions defined on
the time interval $\mathcal{T}$. Denote $\|\cdot\|^2$ as the $L^2$ norm
for functions in $L^2(\mathcal{T})$, that is, $\|f\|^2 \,\hat{=}\,\int_{\mathcal{T}}\{f(t)\}^2\,dt$, $\forall f(t)\in L^2(\mathcal{T})$. The
inner product of two functions $f_1(t)$ and $f_2(t)$ in $L^2(\mathcal
{T})$ is defined as $\langle f_1,f_2\rangle\,\hat{=}\,\int_{\mathcal
{T}} f_1(t)f_2(t)\,dt$. When $\langle f_1,f_2\rangle=0$, we say that
$f_1(t)$ and $f_2(t)$ are orthogonal to each other, denoted as
$f_1\perp f_2$.

\subsection{Regression based principal component analysis for growth
data}\label{PCAsparse}
Reference growth data can be considered as sparse functional data due
to the sparse and irregular data structure. There exists a few PCA
methods for sparse functional data, including \citet{Yao2005}, \citet
{James2000} and \citet{peng2009}. \citet{Yao2005} involved the estimation
of high-dimensional covariance matrices, as well as their inverses,
which may not be computationally stable. \citet{James2000} provided a
stable maximum likelihood estimation (MLE) algorithm under the
assumption of Gaussian process. \citet{peng2009} implemented the same
model from \citet{James2000} using an improved fitting procedure.
However, the distribution assumption of the MLE methods may not be
satisfied by reference growth data. In this section, we propose a
regression based PCA algorithm which is computationally stable, not
relying on strong distribution assumptions, and easily incorporates
covariates. Without loss of generality, and to simplify the notation,
we assume in this section that the population mean $U(t)$ in (\ref
{Ydecompsition}) is 0. For nonzero $U(t)$, we can get its nonparametric
estimation and subtract it from $Y_i(t)$. The algorithm can be applied
to the remaining part as discussed in Remark~\ref{remarklocation}.

The proposed algorithm is based on the fact in
\citet{graves2009functional} that, given $\phi_l(t)$, $1\leq l<k$, and
$r_{ik}$'s, the $k$th component function $\phi_k(t)$ is the minimizer
of the objection function
%
\begin{equation}
\label{objall}
E\bigl\|Y_i(t)-r_{ik}\phi_k(t)
\bigr\|^2,
\end{equation}
subject to the constraints that $\|\phi_k\|^2=1$ and $\phi_k\perp
\phi
_l,\forall1\leq l<k$. And given $\phi_k(t)$, the component score is
%
\begin{equation}
\label{objall2}
r_{ik}=\langle Y_i,\phi_k
\rangle=\arg\min_{r}\bigl\|Y_i(t)-r
\phi_k(t)\bigr\|^2.
\end{equation}
These optimizations provide a theoretical basis for estimating $\phi
_k(t)$ and $r_{ik}$ iteratively and sequentially.

Naturally, a sample version of the objective function (\ref{objall})
can be constructed by
\[
\frac{1}{\sum_{i=1}^Nm_i}\sum_{i=1}^{N}\sum
_{j=1}^{m_i}\bigl|Y_{ij}-r_{ik}
\phi _k(T_{ij})\bigr|^2.
\]
%
Moreover, to estimate $\phi_k(t)$, we approximate it through B-spline
approximations, that is, there exists a $\bolds{\alpha}_k\in\mathbb
{R}^{\ell
_N}$, such that $\phi_k(t)\approx\bolds{\pi}(t)^T\bolds{\alpha
}_k$, where $\bolds{\pi}
(t) =\{
\pi_1(t),\ldots,\pi_{\ell_N}(t)\}^T$ are $\ell_N$ B-spline basis
functions given the specific knots and order. \citet{de1978practical}
showed that any smooth function can always be well approximated by a
B-spline representation with a sufficient number of knots. The
selection of knots and order in practice is discussed in Remark~\ref
{RemarkBspline}. With the above approximations, we have the following
working objective function:
%
\begin{eqnarray}
D_{L^2}(\bolds{\alpha}_k, R_k) &=&
\frac{1}{\sum_{i=1}^N m_i}\sum_{i=1}^N\sum
_{j=1}^{m_i}\bigl|Y_{ij}-r_{ik}\bolds{
\pi}(T_{ij})^T\bolds{\alpha}_{k}\bigr|^2,
\nonumber
\\[-8pt]
\label{equaalphatarget}
\\[-8pt]
\eqntext{\displaystyle\mbox{s.t.}\quad \|\bolds{\pi}(t)^T\bolds{\alpha}_k
\|^2=1\mbox{ and }\bolds{\pi}(t)^T\bolds{\alpha}
_k\perp \bolds{\pi}(t)^T\bolds{\alpha}_l,
\forall1\leq l<k,}
\end{eqnarray}
where $R_k = (r_{1k},\ldots,r_{Nk})^T$ is the vector of the $k$th
component scores.

In what follows, we present a sequential and iterative algorithm to
estimate $\bolds{\alpha}_k$ and $R_k$ in (\ref{equaalphatarget}).
Our proposed
alogrithm is inspired by the iterative least square method in \citet
{WordH1966}, which was used to conduct multivariate PCA. A similar
algorithm in alignment with robust regressions was studied in \citet
{chen2008lower}. However, our algorithm is the first attempt to
implement such an iterative algorithm in PCA for sparse functional data.

\subsubsection*{Estimating the $1$st component}
The\vspace*{1pt} algorithm starts with
estimating the 1st component $(\bolds{\alpha}_1 , R_1)$. We use
$\bolds{\alpha}
_1^{(\nu)}$ and $R_1^{(\nu)}$ for the estimates of $\bolds{\alpha
}_1$ and
$R_1$ at the $\nu$th iteration. The algorithm includes the following steps:
\begin{longlist}[{}]
\item[\textit{Step} 1: \textit{Initial values}.] Generate\vspace*{1pt} $R_1$ with each of its elements
following uniform $(0,1)$ distribution and denote it as $R_1^{(0)}$.

\item[\textit{Step} 2: \textit{Alternating regressions}.]
Continue from the $\nu$th
iteration step with $R_1^{(\nu)}$.
We obtain $\bolds{\alpha}_1^{(\nu+1)}$ by
%
\begin{equation}
\label{alphaupdate}
\bolds{\alpha}_1^{(\nu+1)}=\arg\min_{\bolds{\alpha}\in\mathbb
{R}^{\ell_N}}\frac
{1}{\sum_{i=1}^N m_i}\sum_{i=1}^N
\sum_{j=1}^{m_i}\bigl\llvert
Y_{ij}-r_{i1}^{(\nu)}\bolds{\pi }(T_{ij})^T
\bolds{\alpha} \bigr\rrvert ^2,
\end{equation}
and then standardize $\bolds{\alpha}_1^{(\nu+1)}$ by
$\frac{\bolds{\alpha}_1^{(\nu+1)}}{\sqrt{\|\bolds{\pi
}(t)^T\bolds{\alpha}_1^{(\nu
+1)}\|^2}}$.
The resulting $\bolds{\alpha}_1^{(\nu+1)}$ satisfies
$\|\bolds{\pi}(t)^T\bolds{\alpha}_1^{(\nu+1)}\|^2=1$. Next we
update the
component scores $R_1^{(\nu+1)}$ by
%
\begin{equation}
\label{rregression}
r_{i1}^{(\nu+1)}=\arg\min_{r\in\mathbb{R}}
\sum_{j=1}^{m_i}\bigl\llvert Y_{ij}-r
\bolds{\pi}(T_{ij})^T\bolds{\alpha }_1^{(\nu
+1)}
\bigr\rrvert ^2, \qquad i=1,2,\ldots,N.
\end{equation}
Here (\ref{rregression}) involves $N$ separate regressions. Continue
iterations until the following two conditions are satisfied:
\begin{longlist}[2.]
\item[1.] The differences of $R_1^{(\nu)}$ and $R_1^{(\nu+1)}$, $\bolds
{\alpha}
_1^{(\nu)}$
and $\bolds{\alpha}_1^{(\nu+1)}$ are less than some small value
$\delta_1$
for all their elements;
\item[2.] The change in the
objective function $ D_{L^2}( \bolds{\alpha}_1,R_1)$
between two consecutive iterations does not exceed a small value
$\delta_2$.
\end{longlist}
\item[\textit{Step} 3: \textit{Solutions}.] We\vspace*{1pt} denote the resulting estimates from step
2 as
the $\widehat{\bolds{\alpha}}_1$ and~$\widehat{R}_1$, which are the
estimates
for $\bolds{\alpha}_1$ and $R_1$.
\end{longlist}
It is easy to see that the objective function $ D_{L^2}( \bolds{\alpha}
_1,R_1)$ is monotonically nonincreasing at each iterative step, and the
algorithm will converge to a local minimizer.

\subsubsection*{Estimating the $k$th component with $k>1$}
When we move to the
$k$th component $(\bolds{\alpha}_k , R_k)$ with $k>1$, we need to
solve the
constrained objective function~(\ref{equaalphatarget}). A~numerical
algorithm directly incorporating such constraints is not
straightforward. However, if subtracting $\sum_{l=1}^{k-1}\widehat{r}_{il}\bolds{\pi}(T_{ij})^T\widehat{\bolds{\alpha}}_l$ from
$Y_{ij}$, and denoting
the resulting residuals as $\xi_{ij}^{(k-1)}$, we then have the
following alternative but equivalent objective function:
%
\begin{equation}
\label{alternateobj}
\frac{1}{\sum_{i=1}^N m_i}\sum_{i=1}^N
\sum_{j=1}^{m_i}\bigl\llvert
\xi_{ij}^{(k-1)}-r_{ik}\bolds{\pi }(T_{ij})^T
\bolds{\alpha} _k\bigr\rrvert ^2,
\end{equation}
subject to the only constraint $\|\bolds{\pi}(t)^T\bolds{\alpha}_k\|
=1$. The
equivalence between (\ref{alternateobj}) and~(\ref{equaalphatarget})
comes from the fact that the component function $\phi_k(t)$ is also the
minimizer of $E\|Y_i^{(k-1)}(t)-\langle Y_i^{(k-1)},\phi_k\rangle\phi
_k(t)\|^2$, where $Y_i^{(k-1)}(t)$ is $Y_i(t)-\sum_{l=1}^{k-1}\langle
Y_i,\phi_{l}\rangle\phi_{l}(t)$. The new objective function (\ref{alternateobj}) of $(\bolds{\alpha}_k,R_k)$ is the same in format as
the one
for $(\bolds{\alpha}_1,R_1)$. Therefore, estimating $(\bolds{\alpha
}_k , R_k)$ can be
achieved in a similar fashion as $(\bolds{\alpha}_1 , R_1)$. The only
difference is at each iteration step, we need to orthogonalize $\bolds
{\pi}
^T(t)\alpha_k$ against the previously estimated $\bolds{\pi
}^T(t)\widehat{\alpha}_l,\forall l<k$ to further improve the computational stability.
The numerical details of orthogonalization are provided in Remark~\ref
{RemarkOrth}. When the observations of the growth paths are
sufficiently dense, the orthogonality holds automatically without the
orthogonalization step. The convergence and nonincreasing property also
hold for each $k$. The R program for the proposed algorithm is provided
in the supplemental documents \citet{SuppZhangAOAS2015}.

At last, to determine the number of necessary components $K$, we
propose a model adequacy measure that is an analog of $R^2$ from
\citet{Croux2003}. It measures the total variability explained by the first
$K$ components, that is,
%
\begin{equation}
\label{Rsquare} R^2(K)=1-\frac{\sum_{i=1}^N\sum_{j=1}^{m_i} \{Y_{ij}-\sum_{k=1}^{K}\widehat{r}_{ik}
\bolds{\pi}(T_{ij})^T\widehat{\bolds{\alpha}}_k \}^2}{\sum_{i=1}^N\sum_{j=1}^{m_i}Y_{ij}^2}.
\end{equation}
We stop the estimation algorithm when $ R^2(K)$ is sufficiently large.
The PCA approximation of $Y_i(t)$ can be returned as $\widehat{Y_i(t)}=\sum_{k=1}^K \widehat{r}_{ik}\bolds{\pi}(t)^T\widehat{\bolds{\alpha}}_k$.

\begin{remark}\label{remarklocation}
The above estimation algorithm assumes that $U(t)=0$, hence, one needs
to properly center the growth paths $Y_i(t)$'s before using the
algorithm. We propose to estimate the mean function $U(t)\,\hat{=}\,E\{
Y(t)\}$ using nonparametric methods, such as B-spline smoothing and
local polynomial smoothing, which provide uniform consistent estimators
of the population mean as shown in \citet{hansen2008uniform}, \citet
{de1978practical} and \citet{fan1996local}.\vspace*{1.5pt} Therefore, the algorithm can
be applied to centered data $Y^*_{ij}=Y_{ij}-\widehat{U}(T_{ij})$,
where $\widehat{U}(t)$ is the estimate of $U(t)$. Here $Y^*_{ij}$ are
asymptotically equivalent to the truly centered data as proved in
\citet
{han2010estimating}.
\end{remark}

\begin{remark}\label{remarkstandarize}
In step 2 of our proposed algorithm, we standardize $\bolds{\alpha
}_k^{(\nu
)}$ by $\frac{\bolds{\alpha}_k^{(\nu)}}{\sqrt{\|\bolds{\pi
}(t)^T\bolds{\alpha}_k^{(\nu
)}\|
^2}}$ in each iteration. The standardization step is to meet the
constraint that $\|\bolds{\pi}(t)^T\bolds{\alpha}_k\|^2=1$. It does
not alter the
value of objection function $D_{L^2}(\bolds{\alpha}_k,R_k)$ since
$r_{ik}\bolds{\alpha}_k^T=r_{ik}cc^{-1}\bolds{\alpha}_k^T $ for any
nonzero real
number~$c$.
\end{remark}

\begin{remark}
The proposed algorithm can also be used to obtain singular value
decomposition of
functional data. Let $\mathbf{Y}(t)=\{Y_1(t),\ldots,Y_N(t)\}^T$,
$\mathbf{R}=(R_1,R_2,\ldots)$, and $\Phi(t)=\{\phi_1(t),\phi
_2(t),\ldots
\}$,
then the decomposition (\ref{Ydecompsition}) can be written as
$\mathbf{Y}(t)=\mathbf{R}\Phi(t)$. If we further decompose $\mathbf
{R}=UD$, where $D$ is
a diagonal matrix, we yield the singular value decomposition for
$\mathbf{Y}(t)$, that is, $\mathbf{Y}(t)=UD\Phi(t)$. This step can
be easily
incorporated to the algorithm, but further decompositions of $\mathbf
{R}$ are
out of interest in our context.
\end{remark}

\begin{remark}\label{RemarkOrth}
Let $\mathbf{W}=\int\bolds{\pi}(t)\bolds{\pi}(t)^Tdt$, where $\bolds
{\pi}(t) =\{\pi
_1(t),\ldots
,\pi_{\ell_N}(t)\}^T$ are the given B-spline basis functions. $\mathbf{W}$
is a $\ell_N\times\ell_N$ matrix. Each element of $\mathbf{W}$ is the inner
product of two basis functions, which can be calculated from numerical
integrations. Since $\mathbf{W}$ is a positive-definite matrix, it can be
decomposed as the cross-product of $\mathbf{W}^{{1}/2}$. In this way,
$\bolds{\pi}
(t)^T\bolds{\alpha}_k\perp\bolds{\pi}(t)^T\bolds{\alpha}_l$ is
equivalent to $(\mathbf{W}^{{1}/2}\bolds{\alpha}_k)^T(\mathbf{W}^{{1}/2}\bolds{\alpha}_l)=0$. The
orthogonalization of
$\mathbf{W}^{{1}/2}\bolds{\alpha}_k$ against $\{\mathbf{W}^{{1}/2}\bolds{\alpha}_l\}
_{l=1}^{k-1}$ can be achieved through Gram--Schmidt orthonormalization
from \citet{Trefethen1997}, which projects $\mathbf{W}^{{1}/2}\bolds
{\alpha}_k$
into the orthogonal space spanned by $\{\mathbf{W}^{{1}/2}\bolds{\alpha
}_l\}
_{l=1}^{k-1}$, obtains the projection as $\mathbf{W}^{{1}/2}\bolds
{\alpha}
_k^{\mathrm{proj}}$, and hence has $\bolds{\alpha}_k^{\mathrm{proj}}$
as the
orthogonalized $\bolds{\alpha}_k$. In each of the iterative steps, we
implement such orthogonalization to update $\bolds{\alpha}_k^{(\nu
)}$, which
makes the final solution of $\widehat{\alpha}_k$ satisfy $\bolds{\pi}
(t)^T\widehat{\alpha}_k\perp\bolds{\pi}(t)^T\widehat{\alpha}_l,l<k $.
\end{remark}

\begin{remark}\label{RemarkBspline}
In practice, we choose the knots of B-spline basis functions to be
$q-1$ equally spaced quantiles of pooled time points, that is, $\frac{1}q,\frac{2}q\cdots\frac{q-1}{q}$th quantiles. In this way, the B-spline
basis functions are determined by $q$ and order. Since there are only
two parameters, it is straightforward to choose them by 5-fold
cross-validation using AIC or BIC criterion. Based on our numerical
experience, the results are not sensitive to the exact locations of knots.
\end{remark}

\begin{remark}\label{RemarkTheory}
The proposed algorithm has a lack of consistency of results for the
estimated principal component functions and scores under the sparsity
setting in this paper. A weak asymptotic result for the principal
component functions under restrictive assumptions exists.
\end{remark}

\subsection{The construction of growth charts for growth paths}
\label{growthchartConstruction}
Through\vspace*{1pt}\break the proposed PCA algorithm, we can approximate $Y_i(t)$ as
$\widehat{U}(t)+\break \sum_{k=1}^K \widehat{r}_{ik}\bolds{\pi
}(t)^T\widehat{\bolds{\alpha}}_k$.
Hence, the percentile ranks of $Y_i(t)$ can be identified by estimating
the multivariate quantiles of $(\widehat{r}_{i1},\ldots, \widehat{r}_{iK})$. Multivariate quantiles consider the joint distribution of
components scores and bring additional insights in screening growth
patterns. The individual percentile ranks determined by component
scores enable the comparisons among subjects, which can be useful for
pediatric practice. For example, subject A is at the 95th percentile
and subject B is at the 97th percentile. Using the percentile ranks, a
pediatrician can prioritize the work by examining the health status of
subject B first, since subject B is more likely to have health issues
given its higher percentile rank.

Due to the lack of natural ordering in a multidimensional space, there
is no universally preferred definition of multivariate quantiles, but
various ideas have been developed in the literature. For example,
\citet
{liu1999} and \citet{zuo2000} used multivariate quantile functions
based on the half-space depth functions. Other approaches have been
given by \citet{Parzen1979}, \citet{Abdous1992}, \citet
{Hettmansperger1992}, \citet{Chaudhuri1996}, \citet{Koltchinskii1997},
\citet{Chakraborty2003}, \citet{McDermott2007} and \citet{wei2008}.
\citet{Serfling2002} presented a nice survey of multivariate quantile
functions and outlined the probabilistic properties that a multivariate
quantile function should have.

In our case, the joint distribution of $(\widehat{r}_{i1},\ldots,
\widehat{r}_{iK})$ is unlikely to follow a certain parametric
distribution due to the complexity of sparse functional data.
Therefore, we propose to determine their multivariate quantiles
nonparametrically using \citet{wei2008}, since this method is also
motivated from growth chart problems, and measuring the spatial
``outlyingness'' of an observation relative to a center, which is the
essential part of growth chart studies. \citet{wei2008} converts the
component scores into the polar coordinate system and builds the
quantile contours by nonparametrically regressing the radiuses with
respect to the angles at various quantile levels. Then, by building a
sequence of nested multivariate quantile contours of the $K$ component
scores, our growth chart can be constructed and used to determine the
percentile ranks of growth paths.

Suppose we want to use our constructed growth chart to screen a growth
path of a new subject, including $m_*$ observed measurements, $\{
T_{*j},Y_{*j}\}_{j=1}^{m_*}$. We first obtain its component scores $ \{
r_{*1},\ldots,r_{*K}\}$ by the following least square regression:
%
\begin{equation}
\label{newsubjectLs} \min_{r_1,\ldots,r_K\in\mathbb{R}}\frac{1}{m_{*}} \sum
_{j=1}^{m_{*}}\Biggl\llvert Y_{*j}-
\widehat{U}(T_{*j})-\sum_{k=1}^Kr_k
\widehat{\phi}_k(T_{*j})\Biggr\rrvert ^2,
\end{equation}
where $\widehat{U}(t)$ and $\widehat{\phi}_k(t)$ are estimated from the
reference growth data. By the estimated component scores, this subject
can then be located on the constructed growth chart. If it stays
outside an extreme quantile contour, such as the 0.95th quantile, we
say that its growth path is more unusual than at least 95\% of its
peers, hence it can be singled out for further clinical investigations.

\subsection{Incorporating covariate effects}\label{covariateSec}
Since incorporating subject level information, such as parental
information and ethnicity, can enhance screening performance, we extend
our proposed method to include a covariate $X$. Suppose the reference
growth data consist of $\{(Y_{ij},T_{ij},X_i),i=1,\ldots,N,j=1,\ldots,m_i\}$, where $X_i$ is the covariate of the $i$th subject. We assume
that the measurement $Y_{ij}$ is observed from
\[
Y_{ij}=Y_i(T_{ij},X_i)+
\epsilon_{ij},
\]
where $Y_i(t,x)$ is the underlying growth path for the $i$th subject,
and depends on both age $t$ and covariate $x$. By extending the
Karhunen--Loeve decomposition, we can write
%
\begin{eqnarray}
\label{CovEquModel1} 
Y_i(t,x) &=& U(t,x)+\sum
_{k=1}^{\infty}r_{ik}\phi_k(t,x),\qquad t
\in\mathcal{T},
\end{eqnarray}
where $U(t,x)$ is the mean function, $\phi_k(t,x)$'s are pair-wise
orthogonal component functions, and $r_{ik}$'s are individual component
scores with respect to $\phi_k(t,x)$. Following similar ideas in
Section~\ref{PCAsparse}, we extend the working objective function
(\ref
{equaalphatarget}) as follows:
%
\begin{eqnarray}
\label{CovEquModel2}
 \label{covobj1}
 & \displaystyle D_x(r_{ik},\bolds{
\alpha}_k)   \,\hat{=}\,  \frac{1}{\sum_{i=1}^N
m_i}\sum
_{i=1}^N \sum_{j=1}^{m_i}
\bigl\llvert Y_{ij}-r_{ik}\bolds{\pi}(T_{ij})^T
\bolds {\alpha}_k\bolds{\mu} (X_i)\bigr\rrvert
^2\quad \mbox{s.t.,} &
\\
\label{covcond1}
&\displaystyle \int\bigl\{ \bolds{\pi}(t)^T\bolds{\alpha
}_k\bolds{\mu}(x)\bigr\} ^2\,dt = 1;&
\\
\label{covcond2}
&\displaystyle \int\bigl\{\bolds{\pi}(t)^T\bolds{
\alpha}_k\bolds {\pi}(x)\bigr\} \bigl\{\bolds{\pi} ^T(t)
\bolds{\alpha} _l\bolds{\mu}(x)\bigr\}\,dt = 0
\qquad\forall 1\leq l<k.&
\end{eqnarray}
%

\begin{figure}[b]

\includegraphics{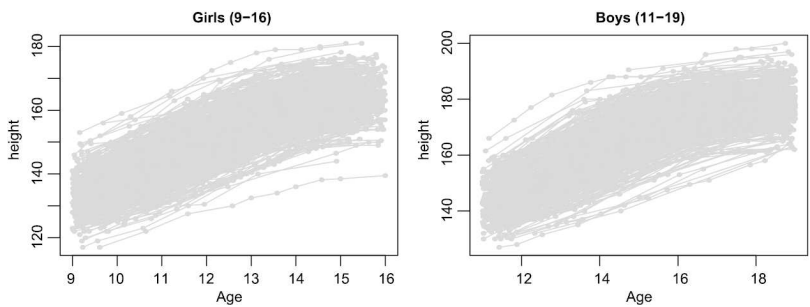}

\caption{Part of a Finnish national growth data from \protect\citet{Pere2000}.
The data include the longitudinal height measurements for 553 girls
(left) from ages 9 to 16 and 518 boys (right) from ages 11 to 19. The
$y$-axis is height and the $x$-axis is age. The dots are the observed
height measurements.}
\label{F001Finnishdata}
\end{figure}

\noindent
Here $\bolds{\pi}(t)^T\bolds{\alpha}_k\bolds{\mu}(x)$ provides
the approximation of $\phi
_k(t,x)$, where $\bolds{\pi}(t)$ is the B-spline basis functions for
$t$ as in
Section~\ref{PCAsparse}, $\bolds{\mu}(x)=\{\mu_{1}(x),\ldots,\mu
_{\ell
_x}(x)\}
^T$ is a set of covariate functions, and $\bolds{\alpha}_k$ becomes a
$\ell
_N\times\ell_x$ matrix instead of a vector. The simplest choice of
covariate functions $\bolds{\mu}(x)$ is $(1,x)^T$, which implicitly assumes
the component functions are linear in $x$ for any given $t$. If the
linearity assumption does not hold, one could consider including
quadratic terms of $x$ or even choosing $\bolds{\mu}(x)$ as B-spline basis\vspace*{1pt}
functions to avoid any parametric assumption. Since $\bolds{\pi
}(t)^T\bolds{\alpha}
_k\bolds{\mu}(x)$ is still a linear function of $\bolds{\alpha}_k$,
we can implement
the similar iterative algorithm in Section~\ref{PCAsparse} by
alternatively updating $\bolds{\alpha}_k$ and $r_{ik}$. The major differences
in each iteration come from the standardization and orthogonalization
of $\bolds{\pi}(t)^T\bolds{\alpha}_k\bolds{\mu}(x)$ in order to
meet constraints (\ref{covcond1}) and~(\ref{covcond2}), details of which are provided in
\citet
{zhang2012regression}. Similarly, the covariate adjusted algorithm is
conducted sequentially, and stopped when reaching an appropriate number
of components\vspace*{1pt} $K$, which is determined by the extended $R^2$, that is,
$1-\frac{\sum_{i=1}^N\sum_{j=1}^{m_i}\{Y_{ij}-\sum_{k=1}^{K}\widehat{r}_{ik}\bolds{\pi}(T_{ij})^T\widehat{\bolds{\alpha}}_k\bolds{\mu
}(X_i)\}^2}{\sum_{i=1}^N\sum_{j=1}^{m_i}(Y_{ij})^2}$.
Then the underlying growth path $Y_i(t,X_i)$ can be well approximated
by the first several component functions, and hence determined by its
component scores. Therefore, the growth chart for screening growth path
can be constructed and implemented in a similar fashion as the one
described in Section~\ref{growthchartConstruction}.

\section{Application examples}\label{ChapterApp}

\subsection{Growth charts for screening pubertal growth paths}\label{secApp}

\begin{figure}[t]

\includegraphics{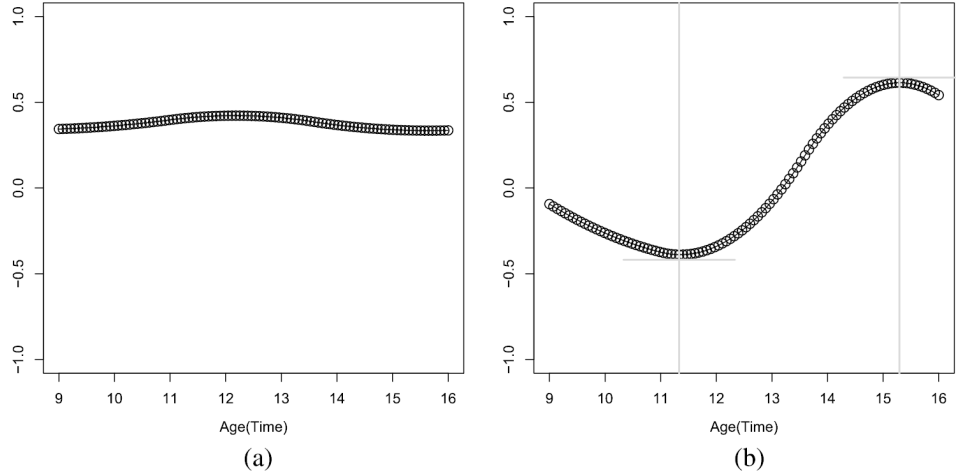}

\caption{The estimated first two component functions $\widehat{\phi}_1(t)$ \textup{(a)} and
$\widehat{\phi}_2(t)$ \textup{(b)} for girls.}
\label{F002girlComponentFunction}
\end{figure}

\begin{figure}[b]

\includegraphics{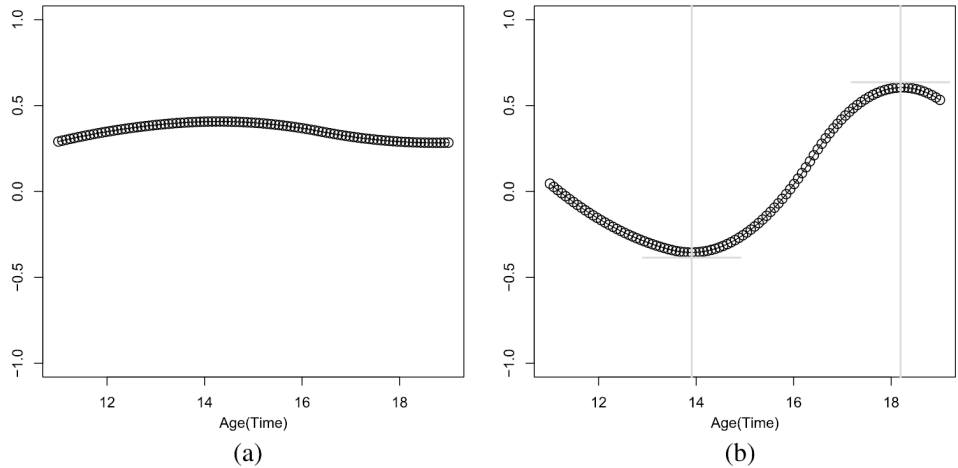}

\caption{The estimated first two component functions
$\widehat{\phi}_1(t)$ \textup{(a)} and $\widehat{\phi}_2(t)$ \textup{(b)} for boys.}
\label{F003boyComponentFunction}
\end{figure}

In this section we illustrate our proposed screening method using part
of a Finnish national growth data set from \citet{Pere2000}. The data
consist of longitudinal height measures of 553 girls (ages 9--16) and
518 boys (ages 11--19) during puberty, as shown in Figure~\ref{F001Finnishdata}. The median number of measurements for each subject
is 6. The analysis is stratified by gender. We apply the proposed
regression based PCA using quadratic B-splines with internal knots 11
and 13.56. The resulting first two component functions are plotted in
Figure~\ref{F002girlComponentFunction} for girls and Figure~\ref{F003boyComponentFunction} for boys. In both cases, they count for
90\% variability of the growth paths based on the proposed $R^2$
measure (\ref{Rsquare}).

\begin{figure}[t]

\includegraphics{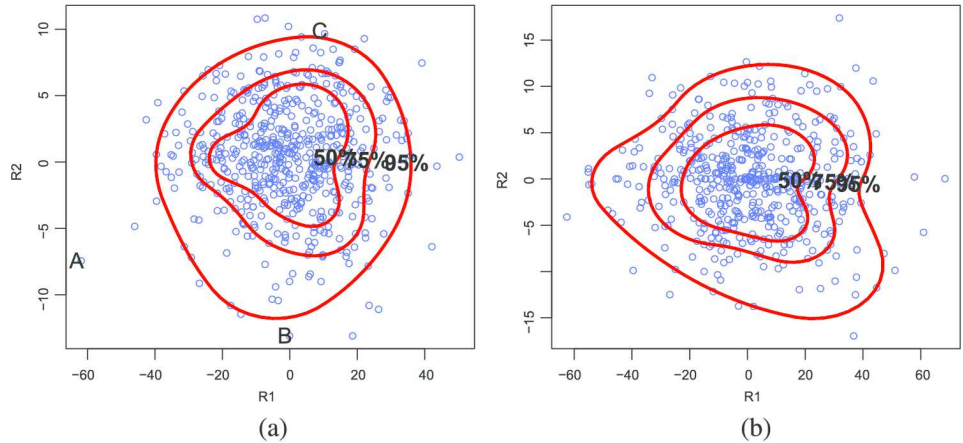}

\caption{The bivariate plot of the first two component scores
for girls \textup{(a)} and boys \textup{(b)}. The $x$ axis represents the first component
score and the $y$ axis represents the second component score. The
contours from inside to outside are the bivariate quantile contours at
quantile levels 0.5, 0.75 and 0.95. The points labeled ``\textup{A}'' and ``\textup{B}'' in
\textup{(a)} are two selected girls whose first two component scores fall
outside the 0.95th quantile contour. \textup{(a)}  The growth chart for girls. \textup{(b)}  The growth chart for boys.}
\label{F004GrowthchartsBoth}
\end{figure}

In both genders, we find that the first component function $\phi_1(t)$
reflects the overall growth scale, while the second one $\phi_2(t)$
coincides well with the puberty growth velocity pattern. The second
component function increases rapidly starting around age 11 and
stabilizes after age 15 for girls [Figure~\ref{F002girlComponentFunction}(b)], while a similar patten is found between
age 14 and age 18 for boys [Figure~\ref{F002girlComponentFunction}(b)].
This difference in $\phi_2(t)$ is biologically reasonable since the
puberty of boys begins later than girls. Therefore, the corresponding
principal component scores have a nice biological interpretation. A
subject with a higher $r_{i1}$ tends to be taller than most of his or
her peers, while a subject with a higher $r_{i2}$ may experience rapid
pubertal growth. The growth charts are constructed based on the first
two component scores, as shown in Figure~\ref{F004GrowthchartsBoth}(a)
for girls and Figure~\ref{F004GrowthchartsBoth}(b) for boys. Such
charts provide a convenient visual tool for screening potentially
unusual growth patterns. In both figures, the $x$ axis represents the
first component score and the $y$ axis represents the second ones.
Bivariate quantile contours at quantile levels 0.5, 0.75 and 0.95 are
added to determine the individual percentile ranks. The individuals
staying outside the 0.95th quantile contour have more outlying
component scores than at least 95\% of their peers. Hence, they will be
screened out for further clinical investigations.

\begin{figure}

\includegraphics{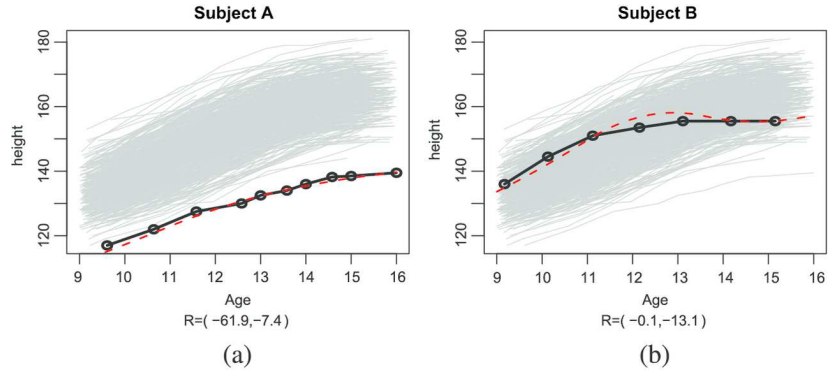}

\caption{The observed growth paths of two extreme girls, girl \textup{A}
\textup{(a)} and girl \textup{B} \textup{(b)} in Figure~\protect\ref{F004GrowthchartsBoth}\textup{(a)}. The black
dots are the original height measurements, and the dashed lines are the
estimated growth paths. The gray background curves are all the growth
paths from the Finnish growth data for girls.}
\label{F005girlOutlyingGrowthPath}
\end{figure}

To illustrate the screening performance of our constructed growth
charts, we select two girls, A and B, who are outside the 0.95th
quantile contour in Figure~\ref{F004GrowthchartsBoth}(a), and further
examine their growth paths as shown in Figure~\ref{F005girlOutlyingGrowthPath}. In Figure~\ref{F005girlOutlyingGrowthPath}, the black dots are the original height
measurements, and the dashed lines are the estimated underlying growth
path $Y_i(t)$. The gray curves in the background are all the growth
paths from the data. According to Figure~\ref{F004GrowthchartsBoth}(a), girl A has small component scores in both
directions, while girl B has an average first component score, but a
very low second component score. Consequently, as shown in Figure~\ref{F004GrowthchartsBoth}, girl A is shorter and slower than most of her
peers; girl B has normative height, but apparently fails to gain enough
height during her puberty. In both cases, the unusual growth patterns
detected by our proposed growth charts are confirmed by empirical
observations of the growth paths.

\begin{figure}

\includegraphics{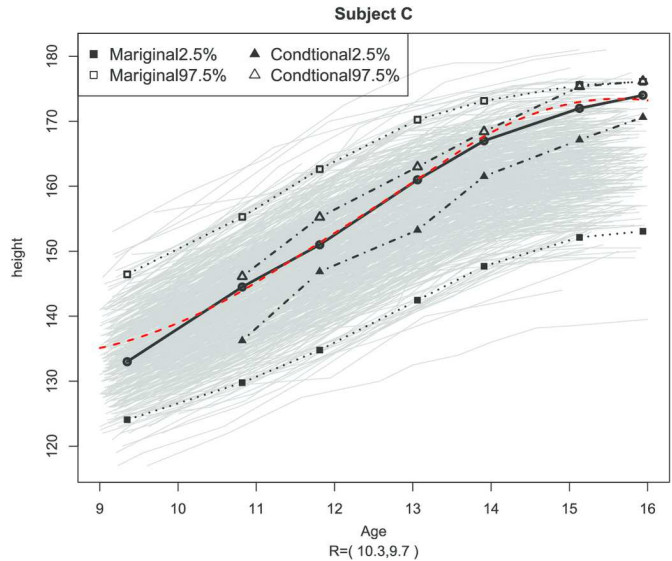}

\caption{The observed growth path of one extreme girl (girl \textup{C}) in
Figure~\protect\ref{F004GrowthchartsBoth}\textup{(a)}. The black dots are the original
height measurements and the dashed line is the estimated growth path.
The gray background curves are all the growth paths from the Finnish
growth data for girls. The squares are the estimated 0.975th (open
squares) and 0.025th (solid squares) quantiles from the unconditional
growth chart. The triangles are the estimated 0.975th (open squares)
and 0.025th (solid squares) quantiles from the conditional growth chart.}
\label{F007ComparewithOtherGrowthcharts}
\end{figure}

\subsubsection*{Comparison to existing growth charts}
As we illustrate in
the \hyperref[SecIntro]{Introduction}, screening entire growth
paths may bring new insights in monitoring human growth. The outlying
girl C in Figure~\ref{F004GrowthchartsBoth}(a) is one example. Figure~\ref{F007ComparewithOtherGrowthcharts} provides the observed growth
path of girl C. Her height starts around the median at the age of 9 and
gradually increases to the upper percentile by the age of 16.

We first screen each of her measurements (black dots) using
conventional growth charts and conditional growth charts. Specifically,
following conventional growth charts from \citet{Wei2006}, we estimate
the 0.025th and 0.975th percentiles that are conditioned only on her
ages (squares in Figure~\ref{F007ComparewithOtherGrowthcharts}). And
following conditional growth charts from \citet{Wei2006}, we estimate
the same reference percentiles conditioned on both her ages and prior
measurements (triangles in Figure~\ref{F007ComparewithOtherGrowthcharts}).

As shown in Figure~\ref{F007ComparewithOtherGrowthcharts}, all of her
height measurements are within the normal ranges of both conventional
and conditional growth charts. Therefore, when these two growth charts
are used to screen her height one at a time, each of her height
measurements is considered as normative. However, when we screen her
entire growth path using the proposed method, girl C is screened out by
the 0.95th quantile contour since her second component score appears
unusually large. It is consistent with the fact that she has been
growing fast consecutively over her entire puberty. This example shows
that the proposed method provides informative insights on growth
pattern by considering entire paths.

\subsection{Growth charts conditioned on mother's height}
\label{SecCovariate}
Parental heights usually have strong associations with their
children's growth. In this section we incorporate mother's height into
the model and examine the pubertal growth of the Finnish teenage girls.
The data set used here is a subset of girls' data in Section~\ref{secApp}, including 444 girls with mother's height information
available and at least 5 measurements between ages 9 and 16. To make
the comparisons, we apply our proposed method, both with covariate and
without covariate, to the data. We choose $\bolds{\mu}(x)$ to be
$\bolds{\mu}(x) =
(1,x)^T$. Under this parameterization, $U(t,x) = U_1(t) + xU_2(t)$,
$\phi_1(t,x) = \phi_{11}(t)+x \phi_{12}(t)$, and $\phi_2(t,x)= \phi
_{21}(t)+x \phi_{22}(t)$. The unknown functions $U_1(t)$, $U_2(t)$,
$\phi_{11}(t)$, $\phi_{12}(t)$, $\phi_{21}(t)$ and $\phi_{22}(t)$ are
all approximated using quadric B-splines with internal knots equal to
$1/3$ and $2/3$ quantiles of pooled times.

\begin{figure}[t]

\includegraphics{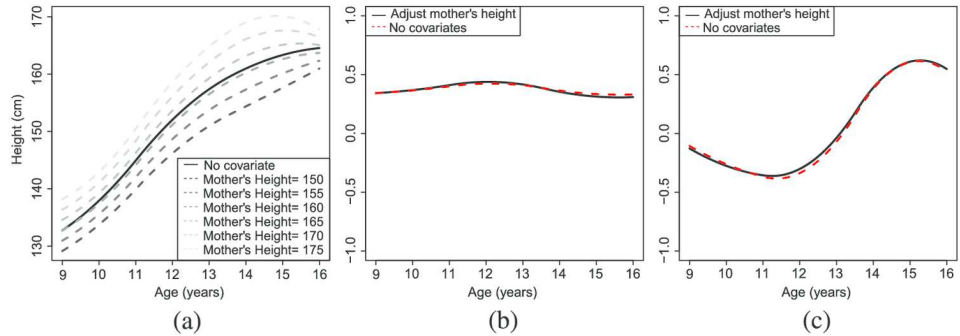}

\caption{\textup{(a)} The estimated mean functions from the covariate adjusted
model (dashed lines) and the model without covariate (solid line). The
dashed lines are the estimated mean functions conditioned on six
different mother's heights. The lines from the lightest gray to the
darkest gray represent 150 cm, 155 cm, 160 cm, 165 cm, 170 cm and 175
cm, respectively. \textup{(b)}, \textup{(c)} The estimated first two component functions from
the covariate adjusted model (dashed lines) and the model without
covariate (solid lines). \textup{(a)} Estimated location functions. \textup{(b)} $\widehat{\phi}_1(t)$  for girls. \textup{(c)} $\widehat{\phi}_2(t)$  for girls.}
\label{F012girlplot52mean}
\end{figure}

\begin{figure}

\includegraphics{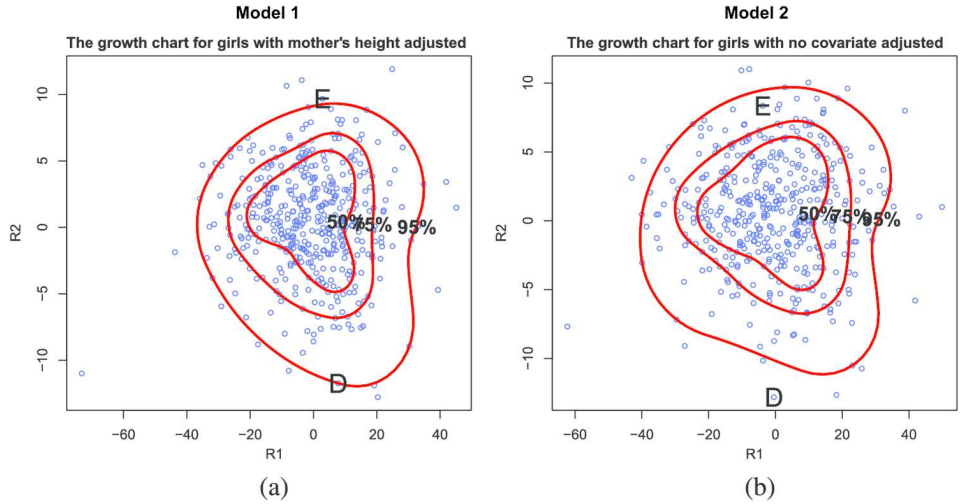}

\caption{The bivariate plots of the first two component scores
for the covariate adjusted model \textup{(a)} and the model without covariate
\textup{(b)}. The $x$ axis represents the first component score and the $y$ axis
represents the second component score. The contours from inside to
outside are the bivariate quantile contours at quantile levels 0.5,
0.75 and 0.95.}
\label{F014girlplot12bivariatecontourX}
\end{figure}

We use a bootstrap to test whether the covariate associated functions
$U_2(t)$, $\phi_{12}(t)$ and $\phi_{22}(t)$ are equal to zero at any
$t$, which is essentially testing whether the corresponding B-spline
coefficients are equal to 0. More details can be found in \citet
{zhang2012regression}. The resulting $p$-values indicate that the
mother's height is significantly related to $U(t,x)$ ($p$-value${}\leq{}$0.0001), while $\phi_1(t,x)$ and $\phi_{2}(t,x)$ are insignificant
($p$-values equal to 0.72 and 0.59). We hence simplify the covariate
adjusted model to
\[
Y_i(t,X_i)\approx U_1(t) +
xU_2(t)+r_{i1}\phi_{11}(t)+r_{i2}
\phi_{21}(t).
\]
In Figure~\ref{F012girlplot52mean}(a), the solid line is the estimated
mean function without considering mother's height, and the dash lines
are the expected growth paths conditioned on six different mother's
heights which are 150 cm, 155 cm, 160 cm, 165 cm, 170~cm and 175 cm
(from darkest gray to the lightest grey), respectively. Covariate
adjusted mean functions show that with the increase of mother's height,
the expected body sizes and growth rates both tend to increase as well.
We also observe the expected growth path conditioned on 160 cm is close
to the expected growth path of the whole population. The explanation is
that the average of mother's height in this data set is 161.6 cm, which
is close to 160 cm. As shown in Figures~\ref{F012girlplot52mean}(b)--(c), the estimated component functions from
both models are very close to each other. However, due to the
difference in the mean functions, the distributions of individual
component scores are fairly different between the two models. Figure~\ref{F014girlplot12bivariatecontourX} plots the bivariate quantile
contours estimated from two sets of component scores. We say that
Figure~\ref{F014girlplot12bivariatecontourX}(a) is the covariate
adjusted growth chart for puberty growth paths and Figure~\ref{F014girlplot12bivariatecontourX}(b) is the marginal one.

\begin{figure}[b]

\includegraphics{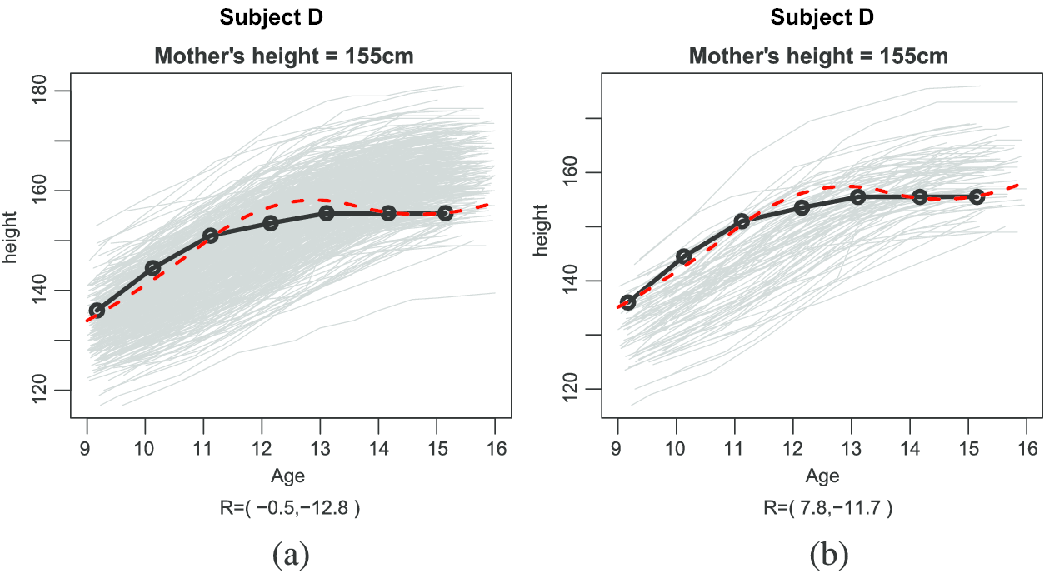}

\caption{The observed growth path of girl \textup{D} in bivariate plots
Figure~\protect\ref{F014girlplot12bivariatecontourX}. The black dots are the original
height measurements. The gray background curves in \textup{(a)} are all the
growth paths from this data set. The gray background curves in \textup{(a)} are
the growth paths of the individuals with mother's height from 153 cm to 155~cm.}
\label{F015girlplot521plotOut}
\end{figure}

\begin{figure}

\includegraphics{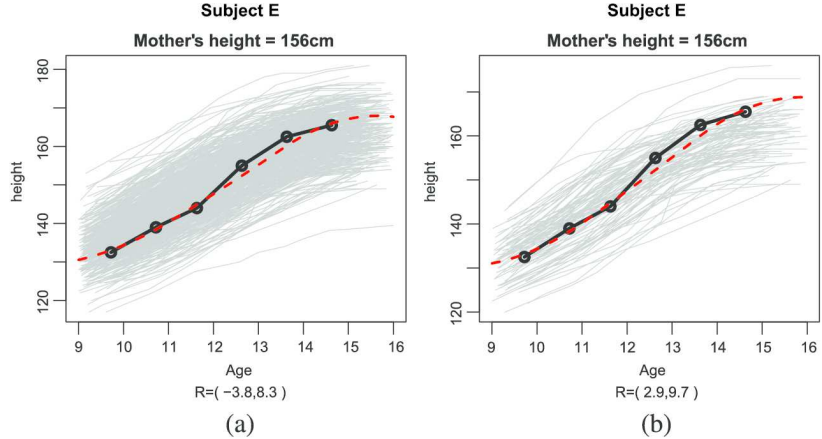}

\caption{The observed growth path of girl \textup{E} in bivariate plots Figure~\protect\ref
{F014girlplot12bivariatecontourX}. The black dots are the original
height measurements. The gray background curves in \textup{(a)} are all the
growth paths from this data set. The gray background curves in \textup{(a)} are
the growth paths of the individuals with mother's height from 154 cm to 158~cm.}
\label{F016girlplot522aplotOut}
\end{figure}

Two girls, D and E, are selected from the sample and placed against
the two growth charts. The growth path of girl D is considered as
unusual in the marginal growth chart, but not in the covariate adjusted
one. In contrast, the growth path of girl E is only considered as
unusual in the covariate adjusted growth chart, but not in the marginal
one. Figures~\ref{F015girlplot521plotOut} and \ref
{F016girlplot522aplotOut} provide their growth paths (black solid
lines and dots) for further investigations. In Figure \ref{F015girlplot521plotOut}(a), we
compare the target paths to all the growth paths in the sample (gray
curves), while in Figures \ref{F015girlplot521plotOut}(b), we compare them only to those (gray
curves) who have similar mother's heights ($\pm$2cm). We find that
girl D has grown unusually slow from ages 12 to 16 compared to others
in the entire sample. That explains why girl D has an unusually low
second component score in the marginal growth chart. However, if one
restricts to those whose mothers have heights around 155 cm, her slow
puberty growth is less extreme, as we observe more similar slow growth
patterns in this subset. Subject E has normative body sizes and growth
rates according to the marginal growth chart, but has excessive growth
based on the covariate adjusted chart. Examining her growth path in
Figure~\ref{F016girlplot522aplotOut}, we find that she has consecutive
years of fast growth from ages 12 to 15. This fast growth appears to be
more extreme when being compared to those whose mothers have similar
heights. In this case, we would have missed the excessive growth of
girl E if we did not take her mother's height into consideration. These
examples show that incorporating subject level information, especially
parental information, might lead to improvements in screening growth paths.

\section{Numerical investigations}\label{ChapterNumerical}
\subsection{Finite sample performance}\label{secNumerical}

In this section we present a numerical simulation study to illustrate
the finite sample performance of the proposed PCA method in comparison
to the alternative \citet{Yao2005} and MLE methods. For MLE methods, we
use the fpca R package based on \citet{peng2009} since it provided an
improved fitting of \citet{James2000}. We consider the following model
to generate the simulation data:
\[
Y_{ij}=Y_i(T_{ij})=U(T_{ij})+r_{i1}
\phi_1(T_{ij})+r_{i2}\phi _2(T_{ij})+
\epsilon_{ij},
\]
where $\phi_1(t)$, $\phi_2(t)$ and $U(t)$ are chosen to be the
estimated functions for girls in Section~\ref{secApp}. We consider the
following two distributions for $(r_{i1},r_{i2})$. In setting 1, we
generate them from the empirical distribution of the estimated first
two component scores for girls in Section~\ref{secApp}. In setting 2,
we generate them from a bivariate normal distribution with sample means
and covariance estimated from the first two component scores for girls
in Section~\ref{secApp}. Both settings try to mimic growth paths of the
Finnish data for girls, while a more restrictive parametric assumption
is made in setting 2. For each of the above two settings, we generate
20 Monte Carlo samples. Each sample includes $N=500$ random curves.
Each one consists of $m_i=6$ observations with the observed time
$T_{ij}$ uniformly distributed on $[9,16]$.

For each sample, we use the proposed method, \citet{Yao2005}, and the
MLE method to conduct PCA. We first estimate $U(t)$ using nonparametric
regression and then apply the three methods to the centered data
$Y_{ij}^*=Y_{ij}-\widehat{U}(T_{ij})$ to estimate component functions.
The selection of tuning parameters for all three algorithms is
described as the following. Because both our method and the MLE method
from \citet{fpcaR} use B-spline functions to represent component
functions, we choose the same set of basis functions for both methods,
that is, the quadratic B-spline basis functions with the $1/3$th and
$2/3$th quantiles of the pooled times as the internal knots. \citet
{Yao2005} relied on estimating the variance and covariance by
two-dimensional local polynomial smoothing. Its smoothing parameters
are determined by minimizing the AIC type criterion, that is, $N\times
\log\{\frac{1}{N}\sum_{i=1}^{N}\frac{1}{m_i}\sum_{j=1}^{m_i}(Y_{ij}-\widehat{Y}_{ij})^2\}+2p$, where $p$ is the number of parameters and
$\widehat{Y}_{ij}$ is the predicted $Y_{ij}$. All codes for the simulations are
written in R language and run under R version 3.0.0 on a machine with
Intel(R) Xeon(R), CPU 3.20 GHz and 16 GB RAM. On average, the running
time to conduct PCA for one Monte Carlo sample is 17 seconds for our
proposed method, 18 seconds for the MLE method, and 30 seconds for
\citet
{Yao2005}.

To evaluate the estimation performance of the three methods, we
calculate relative integrate squares errors (RISE) for both $\phi_1(t)$
and $\phi_2(t)$, where RISE for estimating a target function $g(t)$ is
defined as $\frac{\|g(t)-\widehat{g}(t)\|^2}{\|g(t)\|^2}$, and $\widehat{g}(t)$
is the estimate. RISE can be considered as noise to signal
measurements. The integrations in RISE are evaluated using the left
Riemann sum [\citet{thomas1988calculus}] with the equal partition of the
whole interval into 100 small intervals. Table~\ref{T001RISE1}
provides the summary of RISEs under both settings. As shown in Table~\ref{T001RISE1}, all three methods perform well in estimating
component functions, although \citet{Yao2005} have slightly larger means
and standard deviations.

\begin{table}[t]
\caption{The summary of RISEs for the three sparse
functional PCA methods}\label{T001RISE1}
\begin{tabular*}{\tablewidth}{@{\extracolsep{\fill}}lccc@{}}
\hline
& \multicolumn{3}{c@{}}{\textbf{Means (standard deviations) of RISE}}\\[-4pt]
&\multicolumn{3}{l@{}}{\hrulefill}\\
& \textbf{Yao et al. (\protect\citeyear{Yao2005})} &  \textbf{The MLE method} &  \textbf{The proposed method}\\
\hline
\multicolumn{4}{@{}l@{}}{\textit{Setting} 1: $(r_{i1},r_{i2})\sim{}$Empirical distribution}\\
RISE of $\phi_1(t)$ &0.0061 (0.0017)&0.0003 (0.0003)&0.0004 (0.0005)\\
RISE of $\phi_2(t)$&0.0955 (0.0545)&0.0022 (0.0009)&0.0020 (0.0015)\\[6pt]
\multicolumn{4}{@{}l@{}}{\textit{Setting} 2: $(r_{i1},r_{i2})\sim{}$Bivariate normal
distribution}
\\
RISE of $\phi_1(t)$&0.0052 (0.0018)&0.0003 (0.0003)&0.0004 (0.0003)\\
RISE of $\phi_2(t)$&0.1076 (0.0872)&0.0023 (0.0012)&0.0027 (0.0014)\\
\hline
\end{tabular*}
\end{table}

We further evaluate the estimation errors of component scores $r_{ik}$
among the three methods. For each Monte Carlo sample, we calculate
relative mean square error (RMSE), defined as $\frac{\sum_{i=1}^{N}(r_{ik}-\widehat{r}_{ik})^2/N}{s^2(r_{ik})}$, where
$\widehat{r}_{ik}$ is the estimator of $r_{ik}$ and $s^2(r_{ik})$ is the sample
variance of $r_{ik}$. RMSE measures the fraction of variance
unexplained caused by estimation errors. \citet{Yao2005} involve the
estimation of the individual covariance matrix and its inverse, which
can be singular or close to singular. When it happens, it can deviate
the estimation of component scores $r_{ik}$. To make a fair comparison,
we exclude the top 5\% extreme square errors in the calculation of
RMSE for \citet{Yao2005}. RMSEs under both settings are summarized in
Table~\ref{T002relativeSquareErrors}. All three methods work well for
the 1st component with average RMSEs less than 5\%. For the 2nd
component scores, the average RMSEs for both our proposed method and
the MLE method increase but still less than 20\%, while the RMSEs for
\citet{Yao2005} tend to be slightly larger.

\begin{table}[t]
\caption{The summary of relative mean square errors (RMSE)
$\frac{\sum_{i=1}^{N}(r_{ik}-\widehat{r}_{ik})^2/N}{s^2(r_{ik})}$ for
the three sparse functional PCA methods}\label{T002relativeSquareErrors}
\begin{tabular*}{\tablewidth}{@{\extracolsep{\fill}}lccc@{}}
\hline
& \multicolumn{3}{c@{}}{\textbf{Means (standard deviations) of RMSE}}\\[-4pt]
&\multicolumn{3}{l@{}}{\hrulefill}\\
& \textbf{Yao et al. (\protect\citeyear{Yao2005})}\tabnoteref{tt1} & \textbf{The MLE method}& \textbf{The proposed method}\\
\hline
\multicolumn{4}{@{}l@{}}{\textit{Setting} 1: $(r_{i1},r_{i2})\sim{}$Empirical
distribution}\\
RMSE of $r_{i1}$ &0.05 (0.04)& 0.01 (0.01)& 0.02 (0.01)\\
RMSE of $r_{i2}$ & 0.69 (0.47) & 0.13 (0.02)& 0.17 (0.03)
\\[6pt]
\multicolumn{4}{@{}l@{}}{\textit{Setting} 2: $(r_{i1},r_{i2})\sim{}$Bivariate normal
distribution}\\
RMSE of $r_{i1}$&0.07 (0.05)&0.01 (0.01)&0.02 (0.01)\\
RMSE of $r_{i2}$&0.87 (0.63)&0.14 (0.03)&0.17 (0.04)\\
\hline
\end{tabular*}
\tabnotetext{tt1}{Note: Yao et al. (\protect\citeyear{Yao2005}) involve the estimation of the
individual covariance matrix and its inverse, which can be singular or
close to singular. When it happens, it can deviate the estimation of
component scores $r_{ik}$. To make a fair comparison, we exclude the
top 5\% extreme square errors in the calculation of RMSE for Yao et
al. (\protect\citeyear{Yao2005}).}
\end{table}

\begin{figure}[b]

\includegraphics{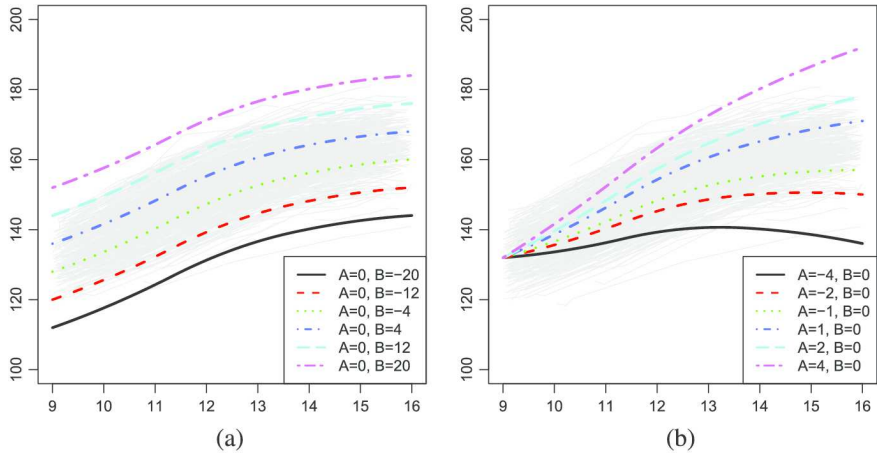}

\caption{\textup{(a)}, \textup{(b)} The selected outlying curves under different combinations of
A and B. The background gray curves are simulated curves from one Monte
Carlo sample.}
\label{F011screenplot}
\end{figure}

\subsection{Screening power}
To illustrate how sensitive the proposed method is in identifying
outlying growth paths compared to the conventional and conditional
growth charts, we simulate Monte Carlo samples from setting 1 as the
reference growth data and build the three types of growth charts
accordingly. We then simulate outlying growth paths $Z_i(t)$ from
$Z_i(t)=Y_i(t)+A(t-9)+B$. Here $Y_i(t)$ follow the correct model from
setting 1, and $A(t-9)+B$ is a linear contaminated term, where $A$
provides the slope deviation and $B$ represents the location shift. We
choose $A$ from $(-4,-2,-1,0,1,2,4)$ and $B$ from
$(-20,-12,-4,0,4,12,20)$. For each $(A,B)$ combination, we generate 100
curves with 6 observations $Z_{ij}=Z_i(T_{ij})+\epsilon_{ij}$ each.
Figure~\ref{F011screenplot} shows the selected outlying curves (dashed
lines) under several combinations of $A$ and $B$. The background gray
curves are from one simulated sample. The simulated curves become more
outlying with the increase of either $|A|$ or $|B|$. Following the
procedure in Section~\ref{growthchartConstruction}, we locate the
simulated outlying curves in the growth charts and screen out those

\section{Conclusion and discussion}\label{conclusion}

This paper develops a new statistical\break method to construct growth charts
for screening entire growth paths. By considering entire growth paths,
the proposed growth charts bring more informative insights into
monitoring pediatric growth. When our constructed growth chart is
applied to the Finnish growth data for monitoring puberty growth, it
shows more effective performance in detecting possible unusual growth
patterns compared to existing growth charts. Besides pediatrics, our
proposed method can also be applied to other areas, such as monitoring
CD4 lymphocyte counts of uninfected children born to HIV-1-infected
women in HIV research, and helping determine the gene frequencies of
the most common mutations in the HFE gene in genetics research.

\begin{sidewaystable}
\tablewidth=\textwidth
\caption{The means of the
percentages of outlying curves $Z_i(t)$ that are screened out by the
95th percentile contours from the proposed growth chart, the 2.5th and
97.5 percentiles from the conventional growth chart, and the 2.5th and
97.5 percentiles from the conditional growth chart for different
combinations of $A$ (slope deviation) and $B$ (location shift)}\label{T002outlyingtable}
\begin{tabular*}{\tablewidth}{@{\extracolsep{\fill}}lccccccc@{}}
\hline
& \multicolumn{7}{c@{}}{\textbf{Means of percentages: The proposed method$\bolds{/}$The
conventional growth chart$\bolds{/}$The conditional growth chart}}\\[-4pt]
& \multicolumn{7}{c@{}}{\hrulefill}\\
&$\bolds{B=-20}$& $\bolds{B=-12}$ & $\bolds{B=-4}$ & $\bolds{B=0}$ & $\bolds{B=4}$ & $\bolds{B=12}$ & $\bolds{B=20}$\\
\hline
$A=-4$ & $100/100/100$ & $98.8/100/100$ & $95.9/98/99.3$ & $93.9/94.3/98.4$ & $91.3/89.2/96.9$ & $86.9/77.1/92.5$ & $87.6/83/84.9$\\
$A=-2$ & $99.4/100/97.2$ & $94.9/96.8/88.9$ & $76.4/73.2/74.5$ & $63.9/51.7/65.6$ & $51.9/33.6/57.2$ & $45/46.4/42.9$ & $67.2/85.2/34.9$\\
$A=-1$& $98/99.2/82.7$ & $77.8/84.2/65$ & $40.2/39.2/45.3$ & $24.2/20.6/36.2$ & $18.9/15.8/28.6$ & $33.7/50/22.1$& $73.8/90.3/24.4$\\
$A=0$& $86.9/95.5/64.6$ & $46.5/62.8/44.3$ & $11.7/15.6/28.6$ & $5.9/9.4/12.8$& $10.8/19.2/22.1$& $47/65/24.6$ & $86.9/95.1/31.4$\\
$A=1$& $69.5/89/64$ & $28.1/48.2/52.1$ & $12.2/14.2/45.5$ & $16.7/22.8/44.5$ & $31.4/43.2/44.7$& $74.4/86.1/48.4$ & $96/98.9/53.4$\\
$A=2$& $60.6/83.7/78.8$ & $37/42.5/76.5$ & $39.5/34.3/75.8$ & $52.1/55/76.4$ & $67.2/75.2/77.9$ & $91.6/97.2/80.7$ & $99/99.7/82.8$\\
$A=4$& $80.3/82.5/96.1$ & $81.3/76.8/96.4$ & $88/90.4/97.7$ & $91.9/95.3/98$ & $95.2/98.3/98.1$ & $99.1/99.8/98.6$ & $100/100/98.7$\\
\hline
\end{tabular*}
\\[12pt]
\tablewidth=\textwidth
\tabcolsep=12pt
\caption{The standard deviations of the
percentages of outlying curves $Z_i(t)$ that are screened out by the
95th percentile contours from the proposed growth chart, the 2.5th and
97.5 percentiles from the conventional growth chart, and the 2.5th and
97.5 percentiles from the conditional growth chart for different
combinations of $A$ (slope deviation) and $B$ (location shift)}\label
{T002outlyingtablesd}
\begin{tabular*}{\tablewidth}{@{\extracolsep{\fill}}lccccccc@{}}
\hline
& \multicolumn{7}{c@{}}{\textbf{Standard deviations of percentages:
The proposed method$\bolds{/}$The conventional growth
chart$\bolds{/}$The conditional growth chart}}\\[-4pt]
& \multicolumn{7}{c@{}}{\hrulefill}\\
&$\bolds{B=-20}$ & $\bolds{B=-12}$ & $\bolds{B=-4}$ & $\bolds{B=0}$ & $\bolds{B=4}$ & $\bolds{B=12}$& $\bolds{B=20}$\\
\hline
$A=-4$&$0.2/0/0$&$1.1/0.2/0.2$ & $2.2/1.4/0.8$ & $3.1/2.6/1.4$ & $3.4/3.5/1.7$ & $3.9/5.2/2.6$ & $3.2/4.4/5.7$\\
$A=-2$ & $0.8/0/2.4$& $2.5/1.8/5.1$&$6.1/7.6/5.8$ & $7.6/7.3/7$ & $7.5/5.9/7.3$ &$7.6/5.9/7.1$&$5.9/4/8.3$\\
$A=-1$ & $1.4/0.8/9.1$ & $6.4/4.5/8.9$ & $7.4/6.7/7.2$ & $7.1/5.4/6.3$ & $7.2/3.6/4.5$ & $7.7/5.7/6.1$ & $6.4/2.9/8.6$\\
$A=0$ & $4.5/1.8/8.6$ & $9.2/6/7.4$ & $4/4.1/4.7$ & $2.8/3.1/4.7$ & $3.7/4.5/4.3$ & $7.7/4.7/7.6$ &$5.7/2.1/11.9$\\
$A=1$ & $12/3/8.8$& $9.8/6.8/8.1$ & $5.4/4/5.3$ & $5.5/4.4/4.6$ & $8.6/4.9/4.4$ & $9.8/4/9.7$ & $3.4/1.2/14.6$\\
$A=2$ & $14.9/3.4/6.9$ & $12.1/7/6$ & $10.3/4.4/4.8$ & $11.8/5.3/4.5$ & $10.9/3.8/5.3$ & $4.3/1.4/7.8$ & $1.3/0.7/10.1$\\
$A=4$ & $6.5/3.8/1.7$ & $5.3/5.6/1.6$ & $4.2/2.6/1.9$ & $3.2/2/1.9$ & $2/1.2/1.9$& $1.1/0.6/1.7$ & $0.2/0/1.7$\\
\hline
\end{tabular*}
\end{sidewaystable}
outside the 95th percentile contours. We also screen each of the
measurements from the simulated outlying curves using the conventional
and conditional growth charts. Specifically, following the conventional
growth chart from \citet{Wei2006}, we estimate the 2.5th and 97.5th
percentiles that are conditioned only on ages. And following the
conditional growth chart from \citet{Wei2006}, we estimate the same
reference percentiles conditioned on both the ages and prior
measurements. Using the conventional and conditional growth charts, we
screen out the curves with more than one measurement outside the range
between the corresponding 2.5th and 97.5th percentiles. Table~\ref{T002outlyingtable} and Table~\ref{T002outlyingtablesd} summarize the
percentages of curves that are screened out by the growth charts,
including both means and standard deviations over 20 Monte Carlo
samples. The results illustrate that all three growth charts are
effective in identifying outlying growth paths when both the location
shift and slope deviation are very extreme ($B=-20$ and $A=-4$). The
conventional growth chart is most sensitive in screening out big
location shifts ($A=0$ and $B=-20,-12,12,20$). The conditional growth chart
works the best for detecting dramatic slope deviations ($B=0$ and
$A=-4,-2,4,2$). The proposed growth chart works the best for identifying
the unusual growth pattern combining moderate location shift and slope
deviation ($B=-4$ and $A=-2$). Among the three growth charts, the proposed
method has the most reasonable type I errors (the results when $A$ and
$B$
are both 0) with mean 5.8\% (9.8\% for the conventional growth chart
and 12.8\% for the conditional growth chart).

The proposed method also contributes to the statistical methodologies.
First, it provides a new way to rank longitudinal/sparse functional
data. It approximates the sparse and irregularly spaced functional data
through PCA and represents each individual using the resulting
components scores. Then the percentile rank of each individual can be
identified by applying multivariate methods to components scores.
Second, the proposed regression based PCA algorithm provides a new way
to conduct PCA for sparse functional data. As shown in Section~\ref{secNumerical}, this algorithm is more computationally stable than
\citet{Yao2005} by avoiding inverting the high-dimensional
variance--covariance matrix. In terms of estimating component functions,
the proposed method is comparable with the MLE method [\citet{peng2009}]. The difference between the proposed method and MLE methods
is essentially the difference between least square regression and MLE
estimator. However, the regression framework has its own advantages
over the likelihood approaches. For example, one can replace the mean
regressions with robust regressions when the data are contaminated with
outliers. In addition, with minor modifications, the proposed
regression based algorithm can also be used to conduct other types of
functional decomposition such as singular value decomposition for
functional data. By supporting various regression models and various
decompositions, the proposed method can be extended to a rich family of
lower dimension approximations for sparse functional data.
Incorporating covariates and conducting variable selections are also
straightforward under the regression framework. Our PCA algorithm
estimates the mean and component functions nonparametrically. If there
are additional recourses indicating certain parametric forms are more
suitable, the efficiency of our method can be further improved.

\begin{supplement}[id=suppA]
\stitle{Supplement to ``Regression based principal component analysis for sparse
functional data with applications to screening growth paths''\\}
\slink[doi]{10.1214/15-AOAS811SUPP} 
\sdatatype{.zip}
\sfilename{AOAS811\_supp.zip}
\sdescription{R programs for the proposed algorithm and an example of
constructing the proposed growth chart.}
\end{supplement}

%

\printaddresses
\end{document}